\documentclass[a4paper,british,conference,12pt]{IEEEtran}
\usepackage[T1]{fontenc}
\usepackage[latin9]{inputenc}
\usepackage{graphicx}

\makeatletter

\pdfpageheight\paperheight
\pdfpagewidth\paperwidth

\providecommand{\tabularnewline}{\\}

\usepackage{url}

\makeatother

\usepackage{babel}
\begin{document}

\title{Channel Decorrelation For Stereo Acoustic Echo Cancellation In High-Quality
Audio Communication}

\author{Jean-Marc Valin\\
CSIRO ICT Centre\\
 Cnr. Vimiera \& Pembroke Roads, Marsfield, NSW, 2122, Australia\\
jean-marc.valin@csiro.au}
\maketitle
\begin{abstract}
In this paper, we address an important problem in high-quality audio
communication systems. Acoustic echo cancellation with stereo signals
is generally an under-determined problem because of the generally
important correlation that exists between the left and right channels.
In this paper, we present a novel method of significantly reducing
that correlation without affecting the audio quality. This method
is perceptually motivated and combines a shaped comb-allpass (SCAL)
filter with the injection of psychoacoustically masked noise. We show
that the proposed method performs significantly better than other
known methods for channel decorrelation.
\end{abstract}

\section{Introduction}

One of the main problems encountered by videoconference users is the
presence of acoustic echo. Sound produced by the loudspeakers is received
by the microphones and sent back to the remote users. This creates
an echo, as users hear the sound of their own voice with a delay.
As videoconferencing applications incorporate higher sampling rates
and multiple channels, the problem of cancelling acoustic echo becomes
harder. One of the main difficulty in stereo echo cancellation is
the strong correlation that exists between the left and right channel,
making it harder or even impossible to estimate the echo filter. 

\begin{figure}
\includegraphics[width=1\columnwidth,keepaspectratio]{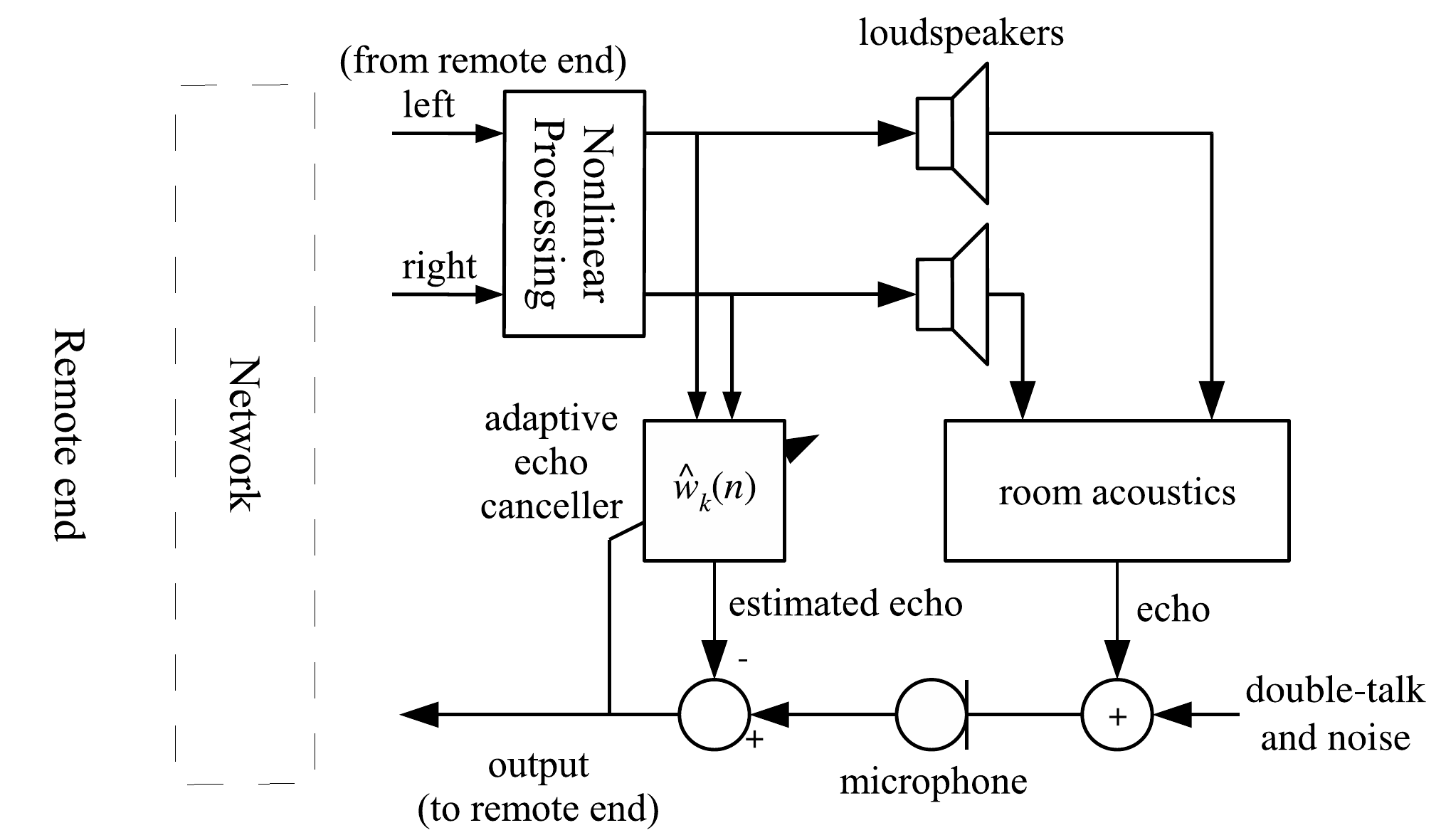}

\caption{Stereo echo cancellation system.\label{fig:Stereo-echo-cancellation} }
\end{figure}

For that reason, it is necessary to reduce the correlation between
channels \cite{Sondhi1995}. This can be done by altering the signals
using some form of non-linear transformation, as illustrated in Fig.
\ref{fig:Stereo-echo-cancellation}. Most of the methods proposed
so far to reduce inter-channel correlation tend to introduce too much
audible distortion to the signal, especially for music. In this paper,
we propose a non-linear processing that closely matches human perception
to maximise decorrelation while minimising the negative impact on
audio quality. 

The paper is divided as follows. Section \ref{sec:Overview-And-Motivations}
presents an overview of the stereo acoustic echo cancellation problem.
Sections \ref{sec:Shaped-Comb-Allpass-Filtering} and \ref{sec:Psychoacoustically-Masked-Noise}
describe the two different parts of the algorithm, respectively the
all-pass filtering and the noise injection. Section \ref{sec:Evaluation-And-Results}
presents the result of comparison with other algorithms and Section
\ref{sec:Conclusion} concludes this paper.

\section{Overview And Motivations\label{sec:Overview-And-Motivations}}

In a multi-channel audio system, there usually exists a correlation
between channels (loudspeakers) that causes the filter optimisation
problem to be under-determined. This means that it can be impossible
to determine the exact contribution of each loudspeaker in the captured
echo because there are an infinity of solutions. It is desirable to
maximise the audio quality, while minimising the inter-channel coherence.
The square coherence is defined as \cite{Morgan2001}:

\begin{equation}
\gamma_{xy}^{2}(f)=\frac{\left|S_{xy}(f)\right|^{2}}{S_{xx}(f)S_{yy}(f)}\label{eq:coherence}
\end{equation}
where $S_{\cdot\cdot}(f)$ denotes the cross-spectrum operator. Assuming
there is no linear transformation involved in the process, the equivalent
frequency-dependent signal-to-noise ratio (SNR) can be expressed as:
\begin{equation}
SNR(f)=\left(\frac{1}{\gamma_{xy}^{2}(f)}-1\right)^{-1}\label{eq:SNR-coherence}
\end{equation}

Many of the approaches proposed so far to reduce inter-channel coherence
have focused on using memoryless non-linearities \cite{Morgan2001}.
The main advantage of memoryless non-linearities is that they are
easy to compute. However, non-linearities of this kind introduce inter-modulation
distortion, which quickly degrades sound quality. Also, there is little
control regarding how much perturbation is caused as a function of
frequency.

Another popular approach is to alter of the phase of the signal in
a time varying way \cite{Ali1998,WU2005}. The time-varying aspect
of the transformation is important because the transformation would
otherwise be linear and thus unable to reduce inter-channel coherence.
The phase of an audio can be altered either through the use of an
all-pass filter, or in the short-term Fourier transform (STFT) domain.

The algorithm we propose in this work was designed to reach the following
goals:
\begin{itemize}
\item Minimising inter-channel coherence
\item Maintaining good quality audio
\item Not altering stereo image in an unpleasant way
\item Not introducing additional delay
\end{itemize}
Because latency is a very important aspect in the perception of acoustic
echo, it is not acceptable for the proposed algorithm to introduce
additional latency. However, it is still possible to use block-based
processing because it is assumed that the transmission and coding
(if any) is performed on blocks. 

\begin{figure}
\begin{center}\includegraphics[width=0.7\columnwidth,keepaspectratio]{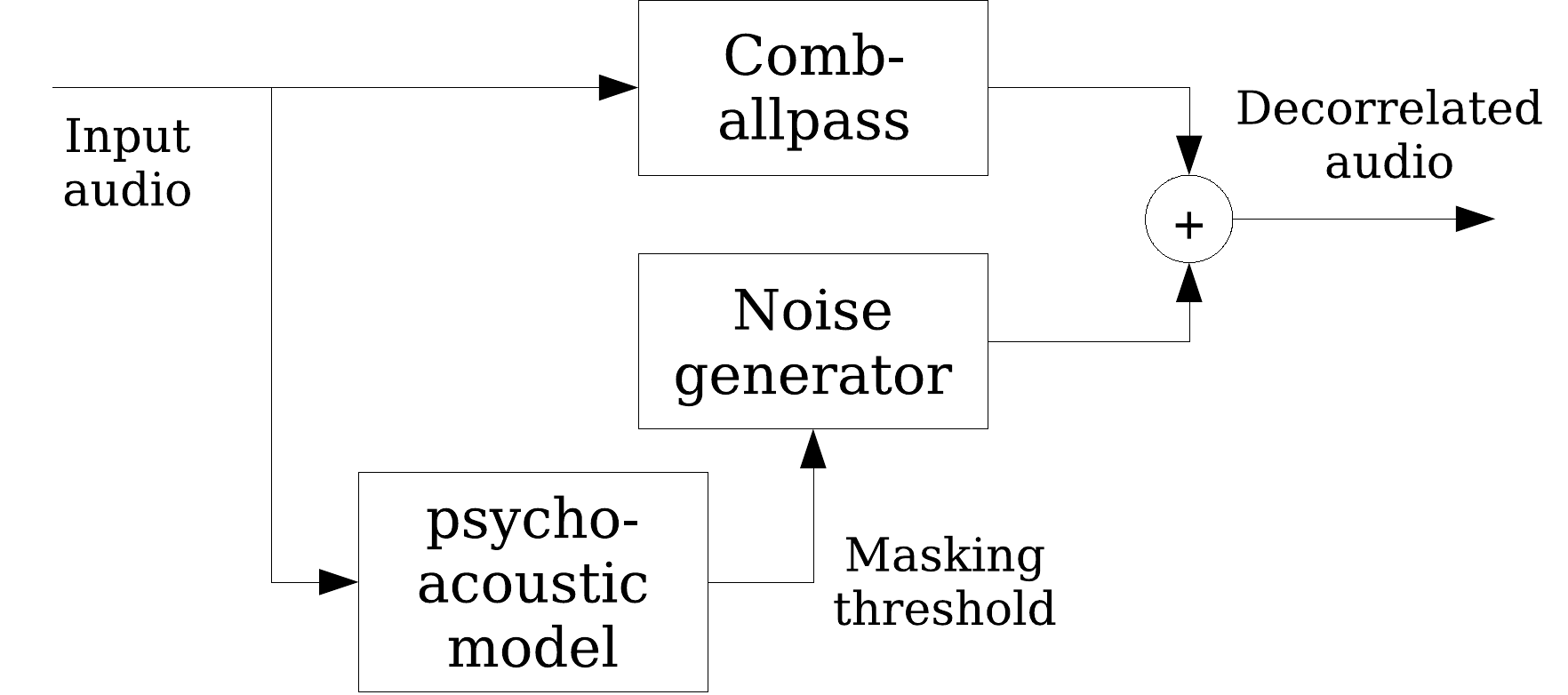}\end{center}

\caption{Overview of the algorithm (for each channel)\label{fig:Overview-of-system}}
\end{figure}

The strategy we propose for decorrelating the audio is illustrated
in Fig. \ref{fig:Overview-of-system} and is divided into two steps:
\begin{itemize}
\item Time-varying phase alteration (mainly at high frequencies)
\item Addition of psychoacoustically-masked noise (mainly at low frequencies)
\end{itemize}

\section{Shaped Comb-Allpass (SCAL) Filtering\label{sec:Shaped-Comb-Allpass-Filtering}}

Allpass filters have a flat frequency response with non-linear phase
and can be represented by the general form:
\begin{equation}
A(z)=\frac{1+\sum_{k=1}^{N}a_{k}z^{k}}{1-\sum_{k=1}^{N}a_{k}z^{-k}}\label{eq:allpass-general}
\end{equation}
The transfer function in Eq. \ref{eq:allpass-general} can be made
causal by adding a constant delay, which leads to:

\begin{equation}
A(z)=\frac{\sum_{k=1}^{N}a_{k}z^{k-N}+z^{-N}}{1-\sum_{k=1}^{N}a_{k}z^{-k}}\label{eq:allpass-causal}
\end{equation}

A filter of the form of Eq. \ref{eq:allpass-causal} is generally
hard to design. However, it is possible to alter the phase similarly
across all frequencies by using a simple comb-allpass filter:

\begin{equation}
A(z)=\frac{\alpha+z^{-N}}{1-\alpha z^{-N}}\label{eq:comb-allpass}
\end{equation}
The filter in Eq. \ref{eq:comb-allpass} combines an all-pole comb
filter to a maximum-phase all-zero comb filter, so the poles and zeros
are equally spread along the frequency axis.

For the processing to be non-linear, it is required to vary the parameter
$\alpha$ controlling the filter. This is achieved through using overlapping
windows with a constant $\alpha$ for each window. We use both an
analysis window and a synthesis window to prevent any blocking artifacts.
The signal is then reconstructed using weighted overlap-add (WOLA).
Because all-pass filtering is a time-domain process, no extra delay
is added because at any given time, we do not need to apply the allpass
filter on whole window. For the analysis-synthesis WOLA process, we
choose the Vorbis window \cite{VorbisSpec}, which meets the Princen-Bradley
criterion \cite{Princen1986} and is defined as:
\begin{equation}
h(n)=\sin\left(\frac{\pi}{2}\sin^{2}\left(\frac{\pi n}{L}\right)\right)\label{eq:Vorbis-window}
\end{equation}

When using a filter of order $N$, there are $N$ points on the unit-circle
where the phase response is zero, regardless of $\alpha$. In other
words, there are frequencies where no decorrelation occurs. For this
reason, it is necessary to also vary the order $N$ of the filter
in so that the ``nulls'' in the phase response change as a function
of time.

\begin{figure}
\includegraphics[width=1\columnwidth,keepaspectratio]{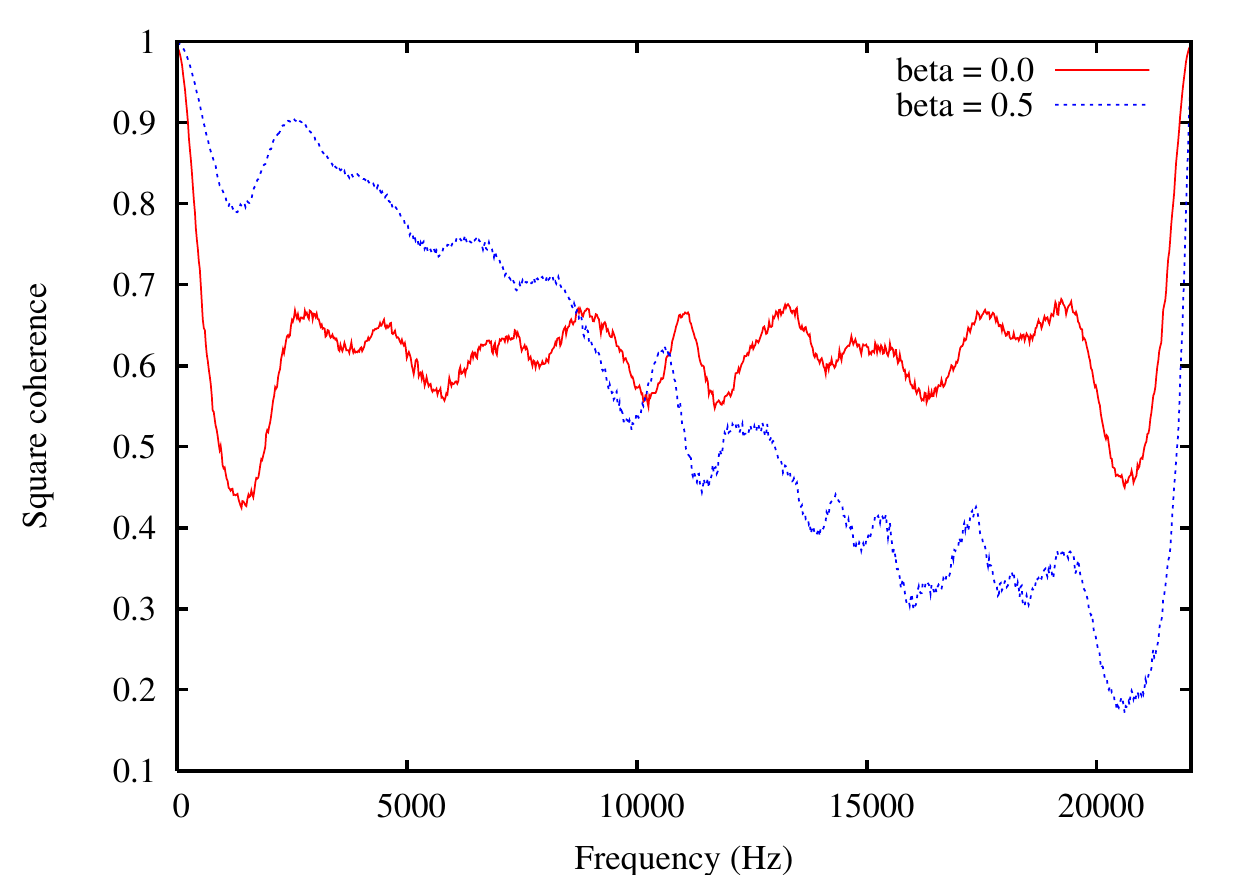}

\caption{Effect of the \emph{tilt} parameter $\beta$ on the square coherence
function for white, Gaussian noise\label{fig:Effect-of-tilt} }
\end{figure}

Interaural phase difference (IPD) is an important localisation cue
at lower frequencies, so the human ear is more sensitive to phase
distortion in the low frequencies. For that reason, it is important
to ``shape'' the phase modulation as a function of frequency. It
is desirable to introduce less distortion to the phase at lower frequencies
than at higher frequencies. To do so, we propose a shaped comb-allpass
(SCAL) filter of the form:
\begin{equation}
A(z)=\frac{\alpha\left(1-\beta z^{-1}\right)+z^{-N}}{1-\alpha\left(-\beta z^{-N+1}+z^{-N}\right)}\label{eq:shaped-comb}
\end{equation}
where $\alpha$ controls the \emph{depth} of the filter and $\beta$
controls the \emph{tilt}. Stability is guaranteed (sufficient condition)
as long as
\begin{equation}
\left|\alpha\right|\left(1+\left|\beta\right|\right)<1\label{eq:stability-criterion}
\end{equation}
The effect of the \emph{tilt} parameter $\beta$ demonstrated in Fig.
\ref{fig:Effect-of-tilt} and can be explained by the fact that as
$\beta$ increases, the poles and zeros of the all-pass filter move
closer to the unit circle at high frequencies and away from the unit
circle at lower frequencies.

For each new window, once the order $N$ is determined,  we choose
$\alpha$ for each frame as:
\begin{equation}
\alpha(N)=\min\left(\left(\alpha(N-1)+r_{0}\right),\frac{1-\epsilon}{1+\left|\beta\right|}\right)\label{eq:random-alpha}
\end{equation}
where $r_{0}$ is a uniformly-distributed random variable chosen in
the $[-r_{max},r_{max}]$ range (typically $r_{max}=0.6$) and $\epsilon\ll1$
controls the distance to the unit circle of the high frequency poles.
The SCAL filter has an overall complexity of only 23 operations per
sample, which is negligible when compared to the complexity of the
adaptive filtering used to cancel the echo.

\section{Psychoacoustically-Masked Noise\label{sec:Psychoacoustically-Masked-Noise}}

The SCAL processing in Section \ref{sec:Shaped-Comb-Allpass-Filtering}
is mainly effective for frequencies above 2 kHz. For lower frequencies,
the ear is more sensitive to phase distortion (altering stereo image),
so it is preferable to inject noise that is uncorrelated to the audio
signal. In this work, we use the psychoacoustic model from the Vorbis
audio codec, as described in \cite{ValinAES2006}. The output of the
psychoacoustic model determines the amount and spectral shape of the
noise that can be added without significantly altering perceptual
audio quality. The psychoacoustic model is also tuned to introduce
less noise in higher frequencies because those are already decorrelated
by the SCAL filter.

The noise to be added is generated in the frequency domain. Again,
we make use of weighted overlap-and-add to reconstruct the time-domain
signal. To avoid adding a delay to the signal of interest, only the
noise is delayed and added to the (non-delayed) input signal. This
is made possible because of the temporal masking effect. The amount
of decorrelation is controlled by the gain $\gamma$ applied to the
noise before adding it to the signal. 

Because we are adding a random signal during the WOLA process, it
is the power that is added and not amplitude. For that reason, we
again need to use a window that satisfies the Princen-Bradley criterion,
and choose the window in Eq. \ref{eq:Vorbis-window}. 

In practice, this step can be made very simple if an audio codec is
already used in the system. If the codec has a low bit-rate, then
it already introduces correlation. Otherwise, it is usually possible
to recover the masking curve from the bit-stream and use it to generate
the noise. This makes it possible to keep the complexity low, again
not contributing significantly to the total complexity of the echo
cancellation process.

\section{Evaluation And Results\label{sec:Evaluation-And-Results}}

In this evaluation, we compare three different decorrelation algorithms:
\begin{itemize}
\item Proposed algorithm
\item Smoothed absolute value
\item First-order, time varying all-pass filter
\end{itemize}
First, the smoothed absolute value non-linearity is defined as:
\begin{equation}
\tilde{x}(n)=x(n)+\alpha\sqrt{x^{2}(n)+c^{2}}\label{eq:smoothed-abs}
\end{equation}
with $c=0.65\sigma_{x}$ as recommended in \cite{Morgan2001}. We
compare with this method because it was shown in \cite{Morgan2001}
to be among the best memoryless non-linearity. The time-varying first
order all-pass filter is implemented as described by \cite{Ali1998}
but using $\alpha_{min}=-.985$ to account for the different sampling
rate used in this work.

The block-based phase alteration method proposed in \cite{WU2005}
was excluded from the comparison because the boundary artifacts caused
by the block processing at high sampling rate causes major quality
degradation, even for very small amounts of decorrelation. While a
WOLA approach could be used, it would involve additional delay, something
which is not acceptable in this context.

In both the proposed algorithm and the first-order all-pass filter,
there is a random component, so it is possible to independently process
each channel with the algorithm. On the other hand, applying the same
memoryless nonlinearity (smoothed absolute value in this case) to
each channel would not reduce the coherence. For that reason, we invert
the sign of the $\alpha$ used for each channel.

\subsection{Methodology}

We evaluate the algorithms by considering the amount of de-correlation
they provide and the degradation in quality they cause. The coherence
in Eq. \ref{eq:coherence} is defined as a function of frequency,
which makes it hard to compare different algorithms. For this reason,
we propose the Bark-weighted square coherence, which we define as:
\begin{equation}
\gamma_{xy}^{2}=\frac{\sum_{f}B'(f)\gamma_{xy}^{2}(f)}{\sum_{f}B'(f)}\label{eq:bark-coherence}
\end{equation}
where $B'(f)$ is the derivative of the Bark scale function $B(f)$,
defined as:
\begin{equation}
B(f)=13\arctan\left(\frac{f}{1316}\right)+3.5\arctan\left(\frac{f^{2}}{7500^{2}}\right)\label{eq:Bark-scale}
\end{equation}
The use of the Bark scale means that each critical band is given the
same weight when computing the coherence value. 

In order to evaluate quality, we use the ITU-R BS.1387 Perceptual
Evaluation of Audio Quality (PEAQ) recommendation \cite{BS1387}.
We use the basic version implementation by Kabal \cite{Kabal2002PEAQ}\footnote{Source code for the software is available from McGill University at:
http://www-mmsp.ece.mcgill.ca/Documents/Software/} with eight different audio excerpts. Six of them (piano, female speech,
male speech, glockenspiel, castanets, quartet) are taken from EBU
Tech 3253 - Sound Quality Assessment Material (SQAM), while the other
two (guitar, Suzanne Vega) are from popular music. We consider the
worst-case scenario of a mono signal going through both loudspeakers,
so the initial coherence is (by definition) equal to 1. All excerpts
are sampled at 44.1 kHz.

The algorithms are tested in six configurations:
\begin{itemize}
\item P1: Proposed algorithm, $\beta=0.62$, $\gamma=0.6$
\item P2: Proposed algorithm, $\beta=0.36$, $\gamma=1.0$
\item P3: Proposed algorithm, $\beta=0.18$, $\gamma=1.67$
\item P4: Smoothed absolute value, $\alpha=0.3$
\item P5: Smoothed absolute value, $\alpha=0.6$
\item P6: First-order all-pass filter, $\alpha_{min}=-0.985$
\end{itemize}

\subsection{Quality and Coherence}

PEAQ quality results in Table \ref{tab:PEAQ-results} show clearly
that P4 and P5 alter quality in an unacceptable (values below -1.5)
way for most samples, while P6 has very unequal quality. Outside of
the obvious results (e.g. P1 outperforms P2, which outperforms P3),
a (paired) Student's t-test reveals with 95\% confidence, that all
configurations of the proposed algorithm (P1, P2, P3) outperforms
all other algorithm configurations (P4, P5, P6). The only exception
is that P3 is considered to outperform P6 with only 80\% confidence. 

\begin{table}
\caption{PEAQ Objective Difference Grade results for all samples. Higher (closer
to 0) is better. Numbers in bold (less than -1.5) are considered to
cause significant (unacceptable) degradation. \label{tab:PEAQ-results}}

\begin{center}%
\begin{tabular}{|c|c|c|c|c|c|c|}
\hline 
Excerpt & P1 & P2 & P3 & P4 & P5 & P6\tabularnewline
\hline 
\hline 
Piano & -0.50 & -1.36 & \textbf{-2.72} & \textbf{-3.14} & \textbf{-3.78} & \textbf{-3.88}\tabularnewline
\hline 
Female speech & -0.31 & -0.51 & -0.83 & \textbf{-2.63} & \textbf{-3.62} & -0.72\tabularnewline
\hline 
Guitar & -0.30 & -0.36 & -0.54 & -0.60 & \textbf{-1.94} & -0.30\tabularnewline
\hline 
Male speech & -0.42 & -0.71 & -1.16 & -1.29 & \textbf{-2.77} & -0.53\tabularnewline
\hline 
Glockenspiel & -0.69 & -1.41 & \textbf{-2.16} & \textbf{-3.88} & \textbf{-3.90} & \textbf{-3.91}\tabularnewline
\hline 
Suzanne Vega & -0.34 & -0.50 & -0.86 & \textbf{-1.59} & \textbf{-3.06} & -0.63\tabularnewline
\hline 
Castanets & -0.59 & -0.62 & -0.86 & \textbf{-3.81} & \textbf{-3.89} & -0.67\tabularnewline
\hline 
Quartet & -0.14 & -0.57 & -1.07 & \textbf{-2.23} & \textbf{-3.53} & \textbf{-1.87}\tabularnewline
\hline 
\end{tabular}\end{center}
\end{table}

Inter-channel coherence for the same samples is shown in Table \ref{tab:Coherence-results}.
Based on these results, we can conclude that P3 provides better decorrelation
(lower coherence) than any other algorithm configuration. P2 is considered
to also decorrelate more than all other algorithms, with the exception
of P5. P1 decorrelates more than P6 (only 80\% confidence) and less
than other configurations. 

\begin{table}
\caption{Bark-weighted square coherence. Lower is better.\label{tab:Coherence-results}}

\begin{center}%
\begin{tabular}{|c|c|c|c|c|c|c|}
\hline 
Excerpt & P1 & P2 & P3 & P4 & P5 & P6\tabularnewline
\hline 
\hline 
Piano & 0.69 & 0.52 & 0.38 & 0.48 & 0.38 & 0.711\tabularnewline
\hline 
Female speech & 0.78 & 0.63 & 0.43 & 0.73 & 0.51 & 0.91\tabularnewline
\hline 
Guitar & 0.78 & 0.63 & 0.42 & 0.73 & 0.36 & 0.92\tabularnewline
\hline 
Male speech & 0.77 & 0.65 & 0.37 & 0.77 & 0.51 & 0.91\tabularnewline
\hline 
Glockenspiel & 0.66 & 0.45 & 0.35 & 0.69 & 0.67 & 0.35\tabularnewline
\hline 
Suzanne Vega & 0.78 & 0.66 & 0.44 & 0.80 & 0.46 & 0.91\tabularnewline
\hline 
Castanets & 0.77 & 0.64 & 0.41 & 0.77 & 0.72 & 0.81\tabularnewline
\hline 
Quartet & 0.79 & 0.63 & 0.44 & 0.64 & 0.39 & 0.86\tabularnewline
\hline 
\end{tabular}\end{center}
\end{table}

\begin{figure}
\includegraphics[width=1\columnwidth]{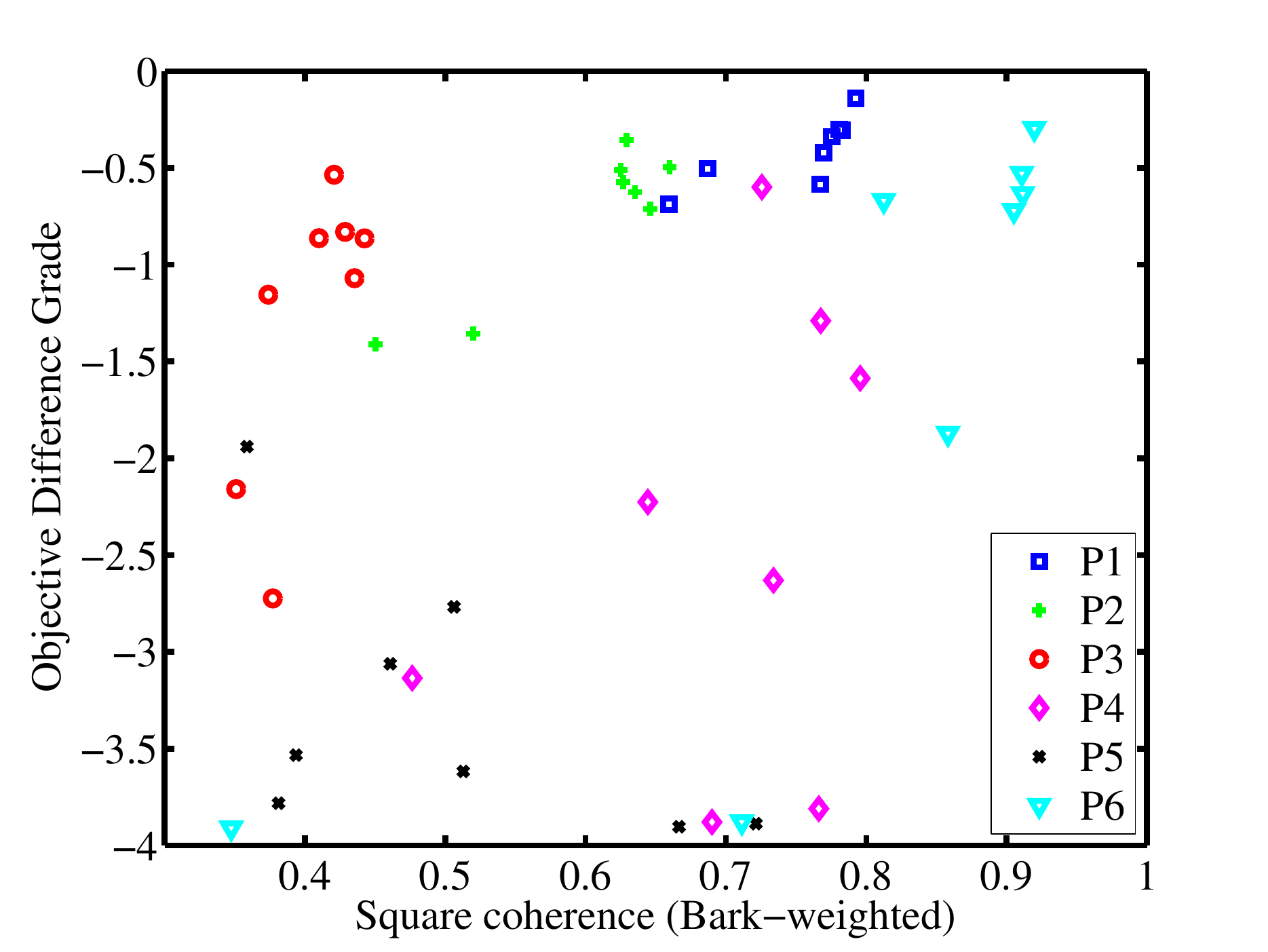}

\caption{Quality as a function of the coherence. Top-left is best, bottom-right
is worst. \label{fig:Scatter-plot}}
\end{figure}

The quality and coherence results are summarised in Fig. \ref{fig:Scatter-plot}.
It can be clearly observed that the proposed algorithm not only performs
better than the other algorithms, but is also much more constant across
all audio excerpts, both in terms of quality and inter-channel coherence. 

Subjective evaluation conducted informally with three listeners based
on the MUltiple Stimuli with Hidden Reference and Anchor (MUSHRA)
\cite{BS1534} methodology are consistent with the PEAQ results and
also show a preference for the proposed algorithm (P1, P2 and P3 ranked
higher than any of the others on average).

\subsection{Analysis of Individual Algorithms}

Listening to the samples makes it possible to identify the various
artifacts caused by the different algorithms and make the following
remarks.

\subsubsection{Proposed algorithm}

\begin{figure}
\includegraphics[width=1\columnwidth,keepaspectratio]{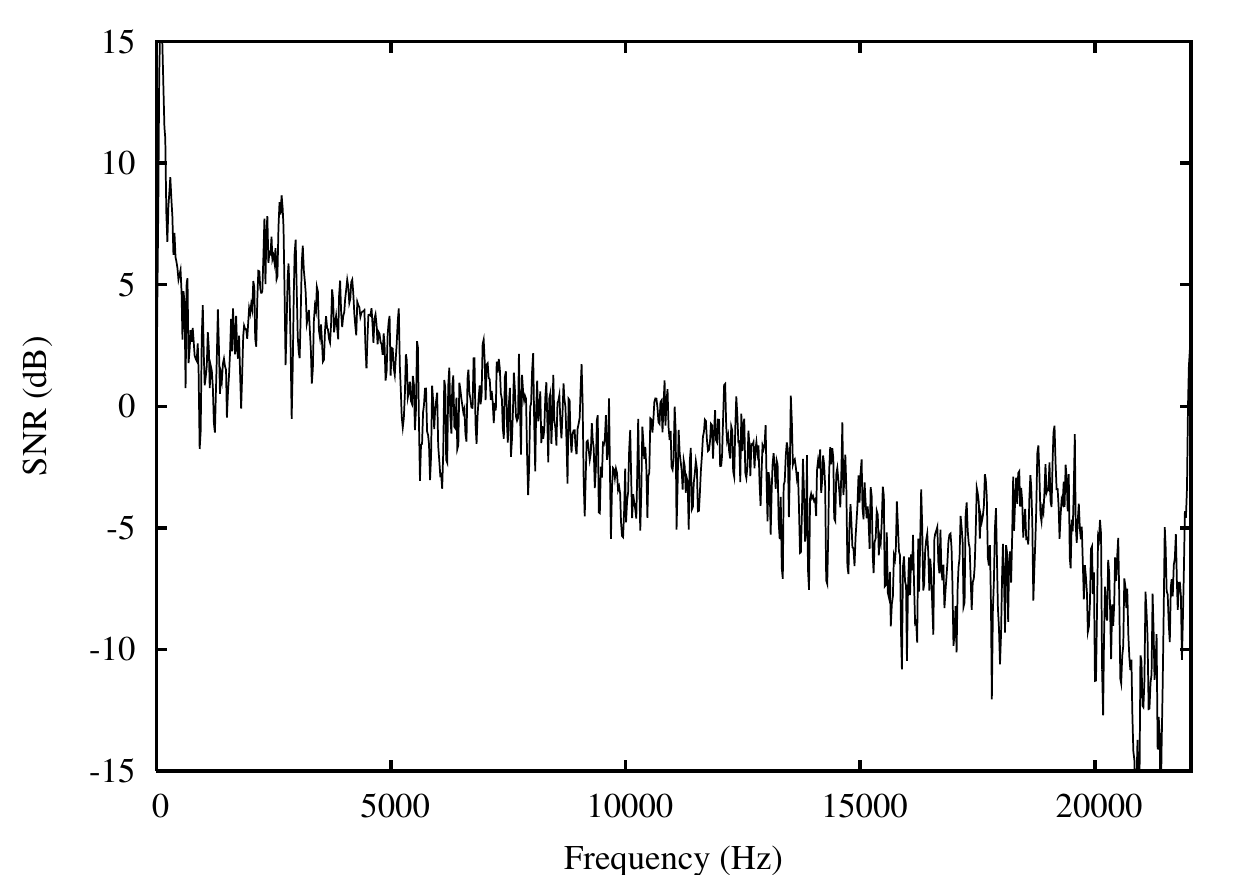}

\caption{Equivalent Signal-to-Noise Ratio (SNR) for P2 on the guitar sample
as a function of frequency.\label{fig:Equivalent-Signal-to-Noise-Ratio}}
\end{figure}

While at low levels (P1), the output of the algorithm is difficult
to distinguish from the original, higher levels (P3) are characterised
by a ``flanging'' effect that is also reflected in the stereo image.
In some samples, some harshness can also be perceived (effect of noise
injection). It can be observed from the equivalent SNR (see Eq. \ref{eq:SNR-coherence})
in Fig. \ref{fig:Equivalent-Signal-to-Noise-Ratio} that the proposed
algorithm is able to introduce a large amount of distortion in the
signal while still maintaining good audio quality (objective difference
grade of -0.36 for the example shown).

\subsubsection{Smoothed absolute value}

\begin{figure}
\includegraphics[width=1\columnwidth]{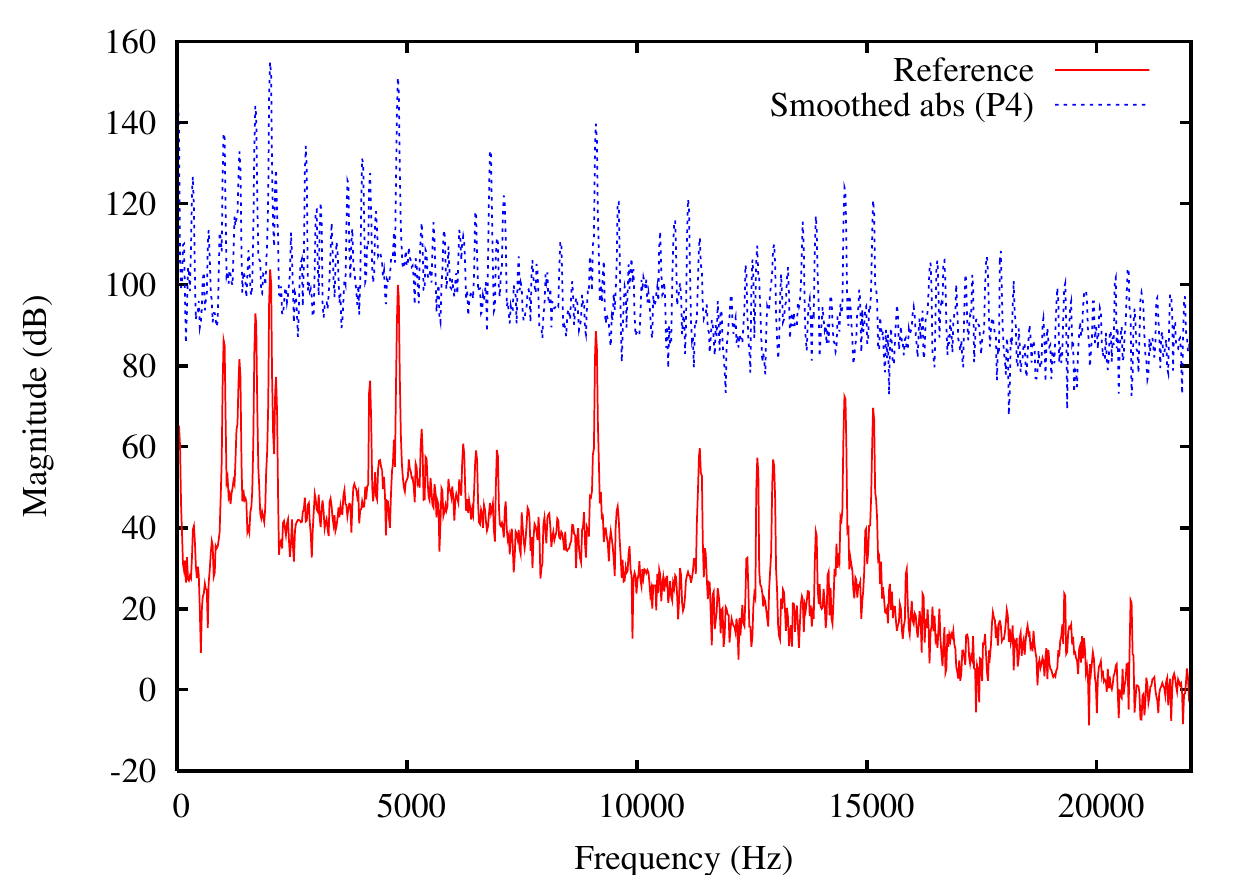}

\caption{Effect of intermodulation distortion on Glockenspiel for smoothed
absolute value (P4). Lower curve: reference, upper curve: distorted.
It can be observed that new tones are created.\label{fig:Distortion}}
\end{figure}

Being a non-linear function, the smoothed absolute value produces
inter-modulation distortion. On harmonic signals such as speech, the
distortion is perceived as additional harshness. However, on tonal
non-harmonic signals such as the glockenspiel, the inter-modulation
distortion effect causes new tones to appear in some regions of the
spectrum. The spectra in Fig. \ref{fig:Distortion} show that the
tones may even appear at low frequencies, which has the effect of
changing the perceived fundamental frequency. 

The last artifact can be observed in the castanets sample. Because
castanets have strong time-domain impulses, the smoothed absolute
value causes one of the channels to be amplified more than the other
depending on the sign of each impulse. This results in a very disturbing
``bouncing'' stereo image, especially for P5.

\subsubsection{First-order all-pass filter}

The main artifact introduced by the first-order all-pass filter is
a nearly white noise that is the result of varying the filter coefficient
$\alpha$ from one sample to another. For most samples, the noise
is masked at lower frequency, so it is usually perceived as a high-frequency
crackling noise. It is mainly perceivable on very tonal samples, that
do not leave much room for masking noise components.

\section{Conclusion\label{sec:Conclusion}}

In this paper, we have demonstrated that it is possible to decorrelate
the left and right channels in a video-conference application without
significantly affecting the audio quality. The proposed method includes
a shaped comb-allpass (SCAL) filter to decorrelate the higher frequencies
and psychoacoustically masked noise injection at lower frequencies. 

The proposed method was shown to outperform other existing methods
both in terms of quality and amount of decorrelation provided. Moreover,
the total complexity of the proposed algorithm is kept small so that
it does not significantly increase the complexity of a complete echo
cancellation system. 

\bibliographystyle{ieeetr}
\bibliography{echo}

\begin{thebibliography}{10}

\bibitem{Sondhi1995}
M.~Sondhi, D.~Morgan, and J.~Hall, ``Stereophonic acoustic echo cancellation-an
  overview of the fundamental problem,'' {\em IEEE Signal Processing Letters},
  vol.~2, no.~8, pp.~148--151, 1995.

\bibitem{Morgan2001}
D.~Morgan, J.~Hall, and J.~Benesty, ``Investigation of several types of
  nonlinearities for use in stereo acoustic echo cancellation,'' {\em IEEE
  Transactions on Speech and Audio Processing}, vol.~9, no.~6, pp.~686--696,
  2001.

\bibitem{Ali1998}
M.~Ali, ``Stereophonic acoustic echo cancellation system using
  time-varyingall-pass filtering for signal decorrelation,'' in {\em
  Proceedings IEEE International Conference on Acoustics, Speech, and Signal
  Processing}, pp.~3689--3692, 1998.

\bibitem{WU2005}
M.~Wu, Z.~Lin, and X.~Qiu, ``A frequency domain nonlinearity for stereo echo
  cancellation,'' {\em IEICE Transactions on Fundamentals of Electronics,
  Communications and Computer Sciences -- Letters}, pp.~1757--1759, 2005.

\bibitem{VorbisSpec}
C.~Montgomery, ``Vorbis {I} specification,'' 2004.
\newblock \url{http://www.xiph.org/vorbis/doc/Vorbis_I_spec.html}.

\bibitem{Princen1986}
J.~Princen and A.~Bradley, ``Analysis/synthesis filter bank design based on
  time domain aliasing cancellation,'' {\em IEEE Transactions on Acoustics,
  Speech, and Signal Processing}, vol.~34, no.~5, pp.~1153 -- 1161, 1986.

\bibitem{ValinAES2006}
J.-M. Valin and C.~Montgomery, ``Improved noise weighting in celp coding of
  speech -- applying the vorbis psychoacoustic model to speex,'' in {\em Proc.
  120th AES Convention}, 2006.

\bibitem{BS1387}
ITU-R, {\em Recommendation BS.1387: Perceptual Evaluation of Audio Quality
  (PEAQ) recommendation}, 1998.

\bibitem{Kabal2002PEAQ}
P.~Kabal, ``An examination and interpretation of {ITU-R BS.1387}: Perceptual
  evaluation of audio quality,'' tech. rep., 2002.
\newblock
  \url{http://www-mmsp.ece.mcgill.ca/Documents/Reports/2002/KabalR2002v2.pdf}.

\bibitem{BS1534}
ITU-R, {\em Recommendation BS.1534-1: Method for the subjective assessment of
  intermediate quality level of coding systems}, 2001.

\end{thebibliography}

\end{document}